\documentclass[bimj,fleqn]{w-art}
\usepackage{times}
\usepackage{w-thm}
\usepackage[authoryear]{natbib}
\theoremstyle{plain}

\theoremstyle{definition}

\usepackage[]{graphicx}
\chardef\bslash=`\\ 

\usepackage{amsmath,amssymb,amsfonts,bm}

\usepackage{booktabs}
\usepackage{multirow}
\usepackage{makecell}
\usepackage{tabularx}
\usepackage{threeparttable}
\usepackage{diagbox}

\usepackage{graphicx}
\usepackage{xcolor}
\usepackage{epstopdf}

\usepackage{algorithm}
\usepackage{algorithmic}

\usepackage{appendix}
\usepackage{lscape}
\usepackage{pdflscape}
\usepackage{enumitem}
\usepackage{multicol}
\usepackage{float}
\usepackage{hyperref}
\usepackage{xr}
\usepackage[T1]{fontenc}
\usepackage{lmodern}
\usepackage{epstopdf}
\usepackage{xcolor}

\newtheorem{assumption}[theorem]{Assumption}

\newtheorem{assumption}{Assumption}
\newtheorem{independentremark}{Remark}

\begin{document}

\DOIsuffix{}
\Volume{}
\Issue{}
\Year{2026}
\pagespan{1}{}


\title[Functional Binomial Model for Competing Risks]{Functional Direct Binomial Regression for Right-Censored Competing Risks Data}

\author[Zheng {\it et al.}]{
Sijie Zheng\footnote{These authors contributed equally to this work.}\inst{,1},
Jonathan H. Hori\footnotemark[1]\inst{,1,2},
Tomoki Okuno\inst{1,2},
Peter Reaven\inst{2},
Greg Norman\inst{3},
Jin J. Zhou\inst{1,2},
Hua Zhou\inst{1,2},
Gang Li\footnote{Corresponding author: e-mail: \textsf{vli@ucla.edu}.}\inst{,1}
}

\address[\inst{1}]{
Department of Biostatistics, University of California, Los Angeles, Los Angeles, CA 90095, USA
}

\address[\inst{2}]{
Phoenix VA Health Care System (111E), Phoenix, AZ 85012, USA
}

\address[\inst{3}]{
Dexcom, Inc., San Diego, CA, USA
}


\begin{abstract}
Existing regression-based methods for functional predictors often overlook right-censored competing risks outcomes, limiting their applicability in many biomedical studies. To address this gap, this article proposes a functional direct binomial model for right-censored competing risks data, where a binary outcome indicates the occurrence of an event of interest before a fixed time horizon. The model enables straightforward prediction of the cumulative incidence function, can be fitted using standard, scalable algorithms for large-scale studies, and has broad applications to functional predictors, such as wearable device data. We establish large-sample theory to facilitate inference, including hypothesis testing and the construction of confidence intervals and bands for both scalar coefficients and functional parameters. The methodology is further extended to accommodate multiple functional biomarkers and to handle missing failure causes. Simulation studies confirm that the proposed method provides accurate estimation and valid inference. We illustrate the approach with an application to modeling time to advanced chronic kidney disease (CKD) in the presence of the competing risk of death, leveraging continuous glucose monitoring (CGM) data and baseline covariates. The analysis shows that CGM is both a significant predictor and a viable alternative to the traditional HbA1c biomarker for CKD.
\end{abstract}

\maketitle

\section{Introduction}

This paper studies competing risks outcomes with right censoring and functional covariates, a setting that arises in many modern applications involving wearable devices, medical
imaging, and electrophysiological signals (e.g., EEG and ECG). Our work is motivated by
risk prediction for long-term complications in diabetes patients using continuous glucose
monitoring (CGM) data, where such complications typically manifest as competing events,
including chronic kidney disease (CKD) and death. Diabetes mellitus (DM) imposes a
rapidly growing global health burden \citep{zhou2024worldwide}. Although CGM devices,
which provide dense measurements of interstitial glucose, have been shown to improve glycemic control and reduce short-term adverse clinical outcomes \citep{beck2017effect, karter2021association,
reaven2023initiation}, it remains unclear whether the rich glucose trajectories captured by CGM can effectively predict long-term complications \citep{battelino2023continuous},
particularly relative to traditional biomarkers such as hemoglobin A1c (HbA1c). This gap is partly due to the lack of analytic tools that can accommodate ultra-dense
longitudinal functional covariates in the presence of right censoring and competing
risks.

A substantial body of work has developed in functional data analysis (FDA),
particularly for linear and generalized linear models with curve-valued covariates
\citep{yao2005functional, hall2007methodology, ramsay2005functional, morris2015functional}.
Foundational studies addressed estimation and inference under dense and sparse designs
\citep{yao2005functional, james2001functional}, followed by advances in dimension reduction
\citep{li2010dimension}, basis expansions \citep{wang2016functional}, and penalized
regression methods \citep{kong2016partially}. More recently, these ideas have been extended to time-to-event settings with right censoring,
leading to functional survival models such as functional Cox regression
\citep{gellar2015cox, lee2015bflcrm, kong2018flcrm, qu2016optimal, hao2021semiparametric}, additive hazards models
\citep{cui2021additive}, accelerated failure time (AFT) models
\citep{liu2025efficient}, censored quantile regression \citep{jiang2020functional},
survival forests \citep{lin2021functional}, causal inference approaches
\citep{gao2024functional}, and joint modeling frameworks for longitudinal and survival
outcomes \citep{li2021flexible, li2022joint}, with a comprehensive review provided by
\citet{li2025some}. 

Despite substantial recent advances in functional survival regression for a single failure time, relatively few methods have been developed for competing risks settings with functional covariates. In competing risks analysis, two primary regression approaches are commonly used: cause-specific hazard regression and regression modeling of the cumulative incidence function (CIF). While existing functional Cox regression methods for a single failure time can, in principle, be applied to cause-specific hazard models, functional covariates have not yet been incorporated into regression frameworks that directly model the CIF, such as the proportional subdistribution hazards model \citep{fine1999proportional} and the direct binomial model \citep{scheike2008predicting}. 

To bridge this gap, this paper introduces a horizon-specific functional direct binomial regression model for the CIF in competing risks data with functional predictors. The model effectively accommodates both scalar and dense functional covariates, naturally addresses the issue of competing risks, and circumvents the assumptions of proportional hazards, as well as the indirect modeling of the CIF. We propose a generalized estimating equation (GEE) method for parameter estimation, utilizing truncated linear basis expansions to reduce the dimensionality of functional covariates, while incorporating inverse probability of censoring weighting (IPCW) to address right censoring \citep{reid1979influence, reid1985influence, lin1989robust}. We establish large-sample theory that supports a diverging number of functional parameters and develop asymptotically valid inference procedures for hypothesis testing, as well as for constructing confidence intervals and simultaneous confidence bands for both scalar coefficients and functional effects. Additionally, we extend our framework to accommodate missing failure causes.
We evaluate the finite-sample performance of the proposed method through extensive simulation studies, assessing estimation accuracy and confidence interval coverage in scenarios with both single and multiple functional covariates. Finally, we apply our approach to model the time to advanced CKD, accounting for the competing risk of death, using CGM data as functional predictors alongside baseline covariates. CGM, as a real-time wearable device, offers significant advantages over traditional biomarkers like HbA1c, which are measured intermittently through blood draws. Our analysis reveals that CGM is a significant predictor of CKD progression, as substantiated by hypothesis testing and confidence band analysis. Furthermore, its predictive performance is comparable to that of HbA1c, suggesting that CGM may serve as a viable alternative biomarker for diabetes management.

The remainder of the paper is organized as follows.
Section~\ref{SEC:model} introduces the proposed model and estimation procedure.
Section~\ref{SEC:thm} develops the asymptotic theory and inference.
Section~\ref{sec:4extensions} discusses several extensions of the proposed framework.
Sections~\ref{SEC:simulation} and~\ref{SEC:app} present simulation studies and a real-data application, respectively.
Section~\ref{sec:discussion} concludes with a discussion.
Technical assumptions, proofs, and additional numerical results are deferred to the Appendix.

\section{Model and Estimation}
\label{SEC:model}
We focus on the standard right-censored competing risks setting with a single functional
covariate. Extensions to settings with multiple functional covariates and missing causes
of failure are discussed in Section~\ref{sec:4extensions}.

\subsection{A Functional Direct Binomial Model for Cumulative Incidence} \label{SEC:2.1}

Let $(T, D)$ be a competing risks outcome, where $T$ denotes the event time and
$D=k$ indicates an event of type $k$, with $1\le k\le K$ ($K\ge 1$). Let
$X(\cdot)\in \mathcal{L}^2[0,1]$ be a functional covariate, 
and let $\boldsymbol{Z} = (Z_1, \ldots, Z_L)^\top$ be an $L$-dimensional vector of time-fixed scalar covariates. 
Without loss of generality, assume that $D=1$
represents the event of interest.

Let $\mu(t)=\mathbb{P}(T\le t, D=1 \mid \boldsymbol{Z}, X(\cdot))$ denote the
conditional CIF for cause 1, and let $t_0$ be a
prespecified prediction horizon. We assume the following horizon-specific functional
direct binomial model:
\begin{equation}\label{eq:model}
\mu(t_0)=g\!\left(\alpha(t_0)+\boldsymbol{\theta}^{\top}(t_0)\boldsymbol{Z}
+\int_{0}^{1}\beta(s;t_0)X(s)\,ds\right),
\end{equation}
where $g(\cdot)$ is a prespecified inverse link function,
$\boldsymbol{\theta}(t_0)=(\theta_1(t_0),\ldots,\theta_L(t_0))^\top$ and
$\beta(\cdot;t_0)$ are the regression coefficients for the scalar and functional
covariates, respectively.

In the absence of functional covariates, model~(\ref{eq:model}) is closely related to
several well-known survival models that directly model the marginal CIF of interest for
competing risks data. For instance, it reduces to the scalar direct binomial model studied
by \citet{scheike2008predicting}. Furthermore, if the model holds for all $t_0 > 0$ with
time-invariant regression coefficients for both the scalar and functional covariates,
it yields the class of transformation models for the CIF considered by
\citet{fine1999analysing} and \citet{mao2017efficient}. This class further includes the
widely used \citet{fine1999proportional} proportional subdistribution hazards model when
the link function is specified as $g^{-1}(u) = \log(-\log(1-u))$ and the proportional odds
model when the link function is $g^{-1}(u) = \log\{u/(1 - u)\}$. To our knowledge, none of these models has been extended to incorporate functional covariates.

\subsection{Estimation}
We now turn to the estimation procedure for the horizon-specific functional binomial model
\eqref{eq:model}, based on right-censored competing risks data consisting of independent
and identically distributed (i.i.d.) observations
$\{(\tilde{T}_i, \delta_i, D_i, \boldsymbol{Z}_i, X_i(\cdot)) : i = 1, \ldots, n\}$,
where $\tilde{T}_i = \min\{T_i, C_i\}$, $\delta_i = I(T_i \le C_i)$, and $C_i$ denotes the
censoring time for subject $i$. 

We assume independent censoring,
$C_i \perp (T_i,D_i,Z_i,X_i(\cdot))$, and positivity,
$\inf_{t\in[0,t_0]}G(t)\ge c_G>0$, where $G(t)=P(C>t)$.
The independent censoring assumption can be relaxed to conditional
independent censoring given the observed covariates; see
Remark~\ref{rmk:1}. Additional regularity conditions for the asymptotic
results are provided in Section~A.2 of the Appendix. For notational simplicity, the dependence on $t_0$ in
model~\eqref{eq:model} is suppressed hereafter.

Assume that $X(\cdot)$ is a square-integrable random function on $[0,1]$ with continuous
mean function $m(s)=\mathbb{E}\{X(s)\}$, and that the regression parameter function
$\beta(\cdot)$ is also square-integrable. Let $\{\rho_j\}$ be an orthonormal basis of
$\mathcal{L}^2[0,1]$.
Then $X(s)$ admits the basis expansion
\(
X(s) = m(s) + \sum_{j=1}^\infty \varepsilon_j \rho_j(s),
\)
where
\(
\varepsilon_j=\int_0^1\{X(s)-m(s)\}\rho_j(s)\,ds
\)
is the score corresponding to the $j$th basis function. In practice, we estimate
$m(s)$ by the sample mean function
\(
\hat m(s)=n^{-1}\sum_{i=1}^n X_i(s)
\)
and compute the score for subject $i$ as
\(
\hat\varepsilon_j^{(i)}
=\int_0^1\{X_i(s)-\hat m(s)\}\rho_j(s)\,ds.
\)
The integrals defining $\hat\varepsilon_j^{(i)}$ are evaluated numerically on the
common grid after the functional trajectories have been reconstructed and aligned.
If the basis functions $\{\rho_j\}$ are chosen as the eigenfunctions of the covariance
operator of $X(\cdot)$, the corresponding scores are the usual functional principal
component scores. In our numerical studies, we use the fixed sine basis
$\rho_j(s)=\sqrt{2}\sin(\pi j s)$.

The proposed framework requires the functional predictors used in the analysis to be represented on a common sufficiently dense grid across subjects, so that the underlying functions and retained basis scores can be adequately represented \citep{muller2005generalized}. The functional predictors may be the raw functional measurements themselves or reconstructed or derived functions obtained from the raw measurements. Thus, the original measurement times may differ across subjects, provided that the resulting functional predictors can be represented on a common dense grid. In our real-data analysis, for example, the functional predictors are derived occupation-time curves indexed by threshold rather than the raw CGM trajectories indexed by calendar time. These curves are represented on a common normalized threshold domain before the basis scores are computed.

Expanding $\beta(s)$ over the same basis yields
\(
\beta(s)=\sum_{j=1}^\infty \beta_j\rho_j(s),
\)
where
\(
\beta_j=\langle\beta,\rho_j\rangle
=\int_0^1 \beta(s)\rho_j(s)\,ds
\)
is the coefficient obtained by projecting $\beta(\cdot)$ onto the $j$th basis function
$\rho_j(\cdot)$,
and $\sum_{j=1}^\infty \beta_j^2<\infty$. By the orthonormality of \( \{\rho_j(\cdot)\} \), model \eqref{eq:model} for subject $i$
can be rewritten as
\begin{align}
\label{eq:beta_0}
g^{-1}(\mu_i)
=\theta_0+\boldsymbol{\theta}^{\top}\boldsymbol{Z}_i
+\sum_{j=1}^\infty \beta_j \varepsilon_j^{(i)},
\end{align}
where $\theta_0=\alpha+\int \beta(s)m(s)\,ds$.

To address the infinite dimensionality of the functional predictor, we adopt the standard
truncation approach \citep{muller2005generalized, kong2018flcrm} and approximate
\eqref{eq:beta_0} by the finite-dimensional model
\begin{align}
\label{eq:error_maintext}
g^{-1}(\mu_i)
=
\theta_0 + \boldsymbol{Z}_i^\top \boldsymbol{\theta}
+ \boldsymbol{\varepsilon}_i^\top \boldsymbol{\beta},
\end{align}
where $\boldsymbol{\beta}=(\beta_1,\ldots,\beta_{p_n})^\top$,
$\boldsymbol{\varepsilon}_i=(\varepsilon_1^{(i)},\ldots,\varepsilon_{p_n}^{(i)})^\top$,
$\boldsymbol{Z}_i=(Z_1^{(i)},\ldots,Z_L^{(i)})^\top$, and $p_n$ is allowed to increase with
the sample size $n$.
This approximation relies on the remainder term
$\sum_{j=p_n+1}^{\infty} \beta_j \varepsilon_j^{(i)}$
being asymptotically negligible as $n \to \infty$, as formalized in
Section~A.1 of the Appendix following
\citet{muller2005generalized}. Proper control of this remainder term is required for
establishing consistency and asymptotic inference under the truncated approximation.

In practice, the truncation level $p_n$ controls the trade-off between
approximation accuracy and estimation complexity. When the effective truncation level
is unknown, we recommend selecting $p_n$ by cross-validation using a prediction
criterion appropriate to the application. In our real-data analysis, we select $p_n$
using cross-validated time-dependent AUC; the detailed selection procedure is provided
in Section~F of the Appendix.

To simplify notation, define
$\boldsymbol{\gamma}=(\theta_0,\boldsymbol{\theta}^\top,\boldsymbol{\beta}^\top)^\top\in\mathbb{R}^{p_n+L+1}$
and
$\boldsymbol{V}_i=(1,\boldsymbol{Z}_i^\top,\boldsymbol{\varepsilon}_i^\top)^\top\in\mathbb{R}^{p_n+L+1}$.
Then \eqref{eq:error_maintext} can be written as
\begin{align}
\label{eq:error_maintext1}
g^{-1}(\mu_i)=\boldsymbol{V}_i^\top\boldsymbol{\gamma}.
\end{align}

Define the binary outcome
$Y_i=I(T_i\le t_0,D_i=1)$, indicating failure from cause 1 before $t_0$.
Then
$\mathbb{E}(Y_i\mid \boldsymbol{Z}_i,X_i(\cdot))=\mu_i$
and
$\mathrm{Var}(Y_i\mid \boldsymbol{Z}_i,X_i(\cdot))=\mu_i(1-\mu_i)$.
We use $\sigma^2(\mu)=\mu(1-\mu)$ to denote the Bernoulli variance function.
In the absence of censoring, $\boldsymbol{\gamma}$ can be estimated by standard binomial
regression, which solves the score equation
\[
\frac{1}{n}\sum_{i=1}^n
\frac{(Y_i-\mu_i)g'(\eta_i)}{\sigma^2(\mu_i)}\boldsymbol{V}_i=0,
\quad \eta_i=g^{-1}(\mu_i).
\]

With right censored competing risks data, however, $Y_i$ and $D_i$ are not observed for all subjects, and the above score equation is no longer applicable. 
In this case, we proposed an approximate generalized estimating equation method based on the IPCW approach, incorporating the competing risks structure and the prediction time $t_0$ \citep{zheng2006application,uno2007evaluating}. Specifically, we estimate $\boldsymbol{\gamma}$ by solving 
\begin{align}
\label{eq:GEE}
\frac{1}{n}\sum_{i=1}^n 
\frac{w_i}{\hat{G} \left(\tilde{T}_i \wedge t_0 -\right)} 
\frac{\left(\tilde{Y}_i-\mu_i\right) g^{\prime}\left(\eta_i\right)}{\sigma^2\left(\mu_i\right)} 
\boldsymbol{V}_i = 0,
\end{align}
where $\tilde{Y}_i = I\left(\tilde{T}_i \leq t_0, \delta_i D_i = 1\right)$ is the observed outcome, 
$w_i = I\left(T_i \wedge t_0 \leq C_i\right) 
= I\left(\tilde{T}_i \leq t_0\right)\delta_i + I\left(\tilde{T}_i > t_0\right)$,  
and $\hat{G}(t)$ is the Kaplan--Meier estimator of the censoring survival function 
$G(t) = \mathbb{P}(C > t)$.  
Note that $\tilde{Y}_i = Y_i$ whenever $w_i = 1$. 
Under independent censoring,
\[
\mathbb{E}\!\left\{
\frac{w_i}{G(\tilde{T}_i \wedge t_0 -)}
\frac{(\tilde{Y}_i-\mu_i) g^{\prime}(\eta_i)}{\sigma^2(\mu_i)}
\boldsymbol{V}_i
\;\middle|\;
\boldsymbol{Z}_i, X_i(\cdot)
\right\}
=
\mathbb{E}\!\left\{
\frac{(Y_i-\mu_i) g^{\prime}(\eta_i)}{\sigma^2(\mu_i)}
\boldsymbol{V}_i
\;\middle|\;
\boldsymbol{Z}_i, X_i(\cdot)
\right\},
\]
so \eqref{eq:GEE} can be viewed as an inverse-probability-weighted
version of the complete-data score equation.
Under the independent censoring and positivity assumptions stated above,
together with the regularity conditions in Section~A.2 of the Appendix,
the resulting estimator is consistent for $\boldsymbol{\gamma}$.

Solving the estimating equation~\eqref{eq:GEE} is computationally straightforward and scalable to large $n$.
However, establishing the asymptotic properties of the resulting estimator is nonstandard,
because (i) the parameter dimension $p_n$ diverges with the sample size, and (ii) the
estimating function involves dependent terms due to the estimation of $\hat{G}$.
A detailed theoretical analysis is provided in the next section and Appendix. For common binary links, the corresponding forms of the score function in
\eqref{eq:GEE} are summarized in Table~S1 in the Appendix B.

\begin{remark}
\label{rmk:1}

When censoring depends on the observed covariates, the IPCW construction
can be extended under conditional independent censoring, where $C_i$ is
independent of $(T_i,D_i)$ conditional on $(Z_i,X_i(\cdot))$. In this case,
the marginal censoring survival function $G(t)$ in~\eqref{eq:GEE} is replaced
by the conditional survival function $G\{t\mid Z,X(\cdot)\}$, together with
a corresponding conditional positivity condition. If
$\widehat G\{t\mid Z,X(\cdot)\}$ is uniformly consistent and admits an
asymptotically linear representation, the same general asymptotic framework
can then be extended under corresponding regularity conditions by replacing
the Kaplan--Meier influence function with that of the conditional censoring
estimator. For example, $\widehat G$ may be obtained from a correctly
specified Cox model for the censoring distribution using the Breslow
estimator, which admits the required asymptotically linear representation
\citep{wolbers2014concordance}.

\end{remark}

\section{Asymptotic Properties and Inference}
\label{SEC:thm}

This section establishes the asymptotic properties of the proposed estimator and develops the associated inferential procedures.
All required assumptions are in Appendix~A.2.

\subsection{Asymptotic Properties}
We first establish the asymptotic properties of the proposed estimators. 
Let $\boldsymbol{\Theta} = \boldsymbol{I}_0 \boldsymbol{\Gamma}_0^{-1} \boldsymbol{I}_0$, where
\[
\boldsymbol{I}_0 = \mathbb{E} \left[ 
\frac{g'^2(\eta)}{\sigma^2(\mu)}
\frac{w}{G^{\dagger}(\tilde{T} \wedge t_0)}
\boldsymbol{V}\boldsymbol{V}^\top
\right], \quad
\boldsymbol{\Gamma}_0 = \mathbb{E} \left[
\frac{g'^2(\eta)}{\sigma^2(\mu)}
\frac{w}{G^{\dagger}(\tilde{T} \wedge t_0)^2}
\boldsymbol{V}\boldsymbol{V}^\top
\right].
\]
Here $G^{\dagger}(t)$ denotes a bias-corrected censoring survival function, defined as
$G^{\dagger}(t)=G(t)^2\{G(t)-\bar{\varphi}(t)\}^{-1}$,
where $G(t)$ is the true censoring survival function and
$\bar{\varphi}(t)=n^{-1}\sum_{i=1}^n \varphi(t;\tilde{T}_i,1-\delta_i)$ captures the first-order
effect of estimating $G(t)$. The explicit form of the influence function
$\varphi(\cdot)$ is given in Appendix~A.2.

\label{SEC:theory}
\begin{thm}\label{thm:beta}
Under Assumptions~A.1--A.5, let
$\hat{\boldsymbol{\gamma}}$ be the solution to~\eqref{eq:GEE}. Then
$$\frac{n\left(\hat{\boldsymbol{\gamma}}-\boldsymbol{\gamma}\right)^{\top} \boldsymbol{\Theta}\left(\hat{\boldsymbol{\gamma}}-\boldsymbol{\gamma}\right)-\left(p_n+L+1\right)}{\sqrt{2\left(p_n+L+1\right)}} \xrightarrow{d} N(0,1) \quad\text{as}\quad n \rightarrow \infty.$$
\end{thm}

The truncation level $p_n$ affects the convergence rate and governs the
approximation error in~\eqref{eq:error_maintext}. As such, $p_n$ acts as a
smoothing parameter that balances estimation variance and approximation bias.
The above asymptotic property holds under the condition
$p_n=o(n^{1/4})$ (Appendix~A.2, Assumption~A.2), which ensures that the
approximation error remains asymptotically negligible relative to estimation
variability. In practice, $p_n$ is selected using a data-driven method; see
Section~\ref{SEC:app} for details.


\begin{remark}
In the absence of censoring, the covariance matrix in
Theorem~\ref{thm:beta} reduces to
$\boldsymbol{\Theta}
=\mathbb E\!\left\{ g'^2(\eta)\sigma^{-2}(\mu)\,
\boldsymbol V\boldsymbol V^\top \right\}$,
which coincides with Theorem~4.1 of \citet{muller2005generalized} when $L=0$.
\end{remark}

To derive the limiting distributions of
$\hat{\boldsymbol{\beta}}$ and $\hat{\boldsymbol{\theta}}$,
write
\(
\boldsymbol{\Theta}^{-1}
=(\boldsymbol{\Theta}_{ij}^{(-1)})_{i,j=1}^2,
\)
where
$\boldsymbol{\Theta}_{11}^{(-1)}\in\mathbb{R}^{(L+1)\times(L+1)}$
and
$\boldsymbol{\Theta}_{22}^{(-1)}\in\mathbb{R}^{p_n\times p_n}$
correspond to the scalar and functional components, respectively.
The following two theorems establish the asymptotic distributions of
$\hat{\boldsymbol{\beta}}$ and $\hat{\boldsymbol{\theta}}$.

\begin{thm}
\label{thm:scb_beta}
Under Assumptions A.1--A.3, A.6--A.8, for $\boldsymbol{\beta}$ and its estimate $\hat{\boldsymbol{\beta}}$,
$$\frac{n\left(\hat{\boldsymbol{\beta}}-\boldsymbol{\beta}\right)^{\top} \left(\boldsymbol{\Theta}_{22}^{(-1)}\right)^{-1}\left(\hat{\boldsymbol{\beta}}-\boldsymbol{\beta}\right)-p_n}{\sqrt{2p_n}} \xrightarrow{d} N(0,1) \quad\text{as}\quad n \rightarrow \infty.$$
\end{thm}

\begin{thm}
\label{thm:scb_theta}
Under Assumptions A.1--A.3 and A.9, for $\tilde{\boldsymbol{\theta}}=\left(\theta_0,\boldsymbol{\theta}^{\top}\right)^{\top}$ and its estimate $\hat{\tilde{\boldsymbol{\theta}}}$,
\[\sqrt{n}\left(\hat{\tilde{\boldsymbol{\theta}}} - \tilde{\boldsymbol{\theta}}\right) \xrightarrow{d} N\left(\mathbf{0}_{L+1},\boldsymbol{\Theta}_{11}^{(-1)}\right) \quad \text{as}\quad n \to \infty.\]
\end{thm}
If the scalar covariates $\boldsymbol{Z}$ and the functional scores
$\boldsymbol{\varepsilon}$ are independent, $\boldsymbol{\Theta}$ is block-diagonal.
Otherwise, both the quadratic form in Theorem~\ref{thm:scb_beta} and the asymptotic
distribution in Theorem~\ref{thm:scb_theta} depend on the full covariate vector
$\boldsymbol{V}$ through $\left(\boldsymbol{\Theta}_{22}^{(-1)}\right)^{-1}$ and
$\boldsymbol{\Theta}_{11}^{(-1)}$.

\subsection{Variance Estimation}
We construct a consistent estimator of the asymptotic covariance matrix.
Define the empirical estimator of $\boldsymbol{\Theta}$ as
$\hat{\boldsymbol{\Theta}}=\hat{\boldsymbol{I}}_{\gamma}\hat{\boldsymbol{\Gamma}}_{\gamma}^{-1}\hat{\boldsymbol{I}}_{\gamma}$,
where
\[
\hat{\boldsymbol{I}}_{\gamma}
=
\frac{1}{n}\sum_{i=1}^{n}
\frac{w_i}{\hat{G}(\tilde{T}_i\wedge t_0-)}
\frac{g'^2(\hat{\eta}_i)}{\sigma^2(\hat{\mu}_i)}
\boldsymbol{V}_i\boldsymbol{V}_i^\top,
\quad
\hat{\boldsymbol{\Gamma}}_{\gamma}
=
\frac{1}{n}\sum_{i=1}^{n}
\frac{w_i}{\hat{G}(\tilde{T}_i\wedge t_0-)^2}
\frac{g'^2(\hat{\eta}_i)}{\sigma^2(\hat{\mu}_i)}
\boldsymbol{V}_i\boldsymbol{V}_i^\top .
\]
Consistency of $\hat{\boldsymbol{\Theta}}$ relies on two key ingredients.
First, additional assumptions are required to control empirical averages
of diverging-dimensional matrices, a task that is substantially more delicate
than in the finite-dimensional setting.
Second, the error induced by replacing the bias-corrected weight
$G^{\dagger}(t)$ with its estimator $\hat G(t)$ is shown to be asymptotically
negligible; see Lemma~A.2 in the Appendix.
Together with Theorem~\ref{thm:beta}, these results establish the
consistency of $\hat{\boldsymbol{\Theta}}$.

Partition $\hat{\boldsymbol{\Theta}}^{-1}$ conformably with $\boldsymbol{\Theta}^{-1}$ as
\(
\hat{\boldsymbol{\Theta}}^{-1}
=(\hat{\boldsymbol{\Theta}}_{ij}^{(-1)})_{i,j=1}^2,
\)
and denote by
$\bigl(\hat{\boldsymbol{\Theta}}_{22}^{(-1)}\bigr)^{-1}$ the inverse of the functional block.

\begin{thm}
\label{thm:variance}
(i) Under Assumptions~A.1--A.5 and A.10,
$\hat{\boldsymbol{\Theta}}\to_p \boldsymbol{\Theta}$.

(ii) Under Assumptions~A.1--A.5, A.10, and A.11,
$\left(\hat{\boldsymbol{\Theta}}_{22}^{(-1)}\right)^{-1}\to_p
\left(\boldsymbol{\Theta}_{22}^{(-1)}\right)^{-1}$.

(iii) Under Assumptions~A.1--A.5, A.10, and A.11(i),
$\hat{\boldsymbol{\Theta}}_{11}^{(-1)}\to_p \boldsymbol{\Theta}_{11}^{(-1)}$.
\end{thm}

\begin{remark}
The covariance estimator $\hat{\boldsymbol{\Theta}}$ is asymptotically positive
definite provided that $\hat{G}(t)>0$.
As with all IPCW-based methods, estimation accuracy may deteriorate when the censoring survival function is close to zero.
In contrast to \citet[][p.~207]{scheike2008predicting}, which require $G(t)$ to be
uniformly bounded away from zero over the entire follow-up period, our analysis
only relies on local positivity of $G^{\dagger}(\tilde T_i\wedge t_0-)$ at the
prediction horizon $t_0$, which suffices for horizon-specific CIF inference.
\end{remark}

\subsection{Inference}
To formally assess the significance of the functional predictor, we consider the
test
\[
H_0:\ \beta(s)=0 \quad \text{vs.} \quad H_a:\ \beta(s)\neq 0 .
\]
Under the truncated representation, this is equivalent to testing
\[
H_0:\ \beta_1=\cdots=\beta_{p_n}=0 
\quad \text{vs.} \quad 
H_a:\ \exists\, 1\le j\le p_n \ \text{such that } \beta_j\neq 0 .
\]

Define the test statistic
\[
T_n=
\frac{
n\hat{\boldsymbol{\beta}}^{\top}
\left(\hat{\boldsymbol{\Theta}}_{22}^{(-1)}\right)^{-1}
\hat{\boldsymbol{\beta}}
-
p_n
}{\sqrt{2p_n}} .
\]
By Theorems~\ref{thm:scb_beta} and~\ref{thm:variance}, $T_n \xrightarrow{d} N(0,1)$ under
$H_0$. A one-sided level-$\alpha$ test rejects $H_0$ when
$|T_n|>z_{1-\alpha/2}$, with $p$-value
$p=2\{1-\Phi(|T_n|)\}$.

Beyond hypothesis testing, we provide uncertainty quantification for the CIF through
asymptotic confidence intervals and bands for the conditional mean $\mu$, the functional
coefficient $\beta(s)$, and the scalar coefficients $\boldsymbol{\theta}$.
Let $\{(e_k,\lambda_k)\}_{k=1}^{p_n+L+1}$ denote the eigenpairs of $\boldsymbol{\Theta}$,
with $e_k=(e_{k,1},\ldots,e_{k,p_n+L+1})^\top$. The following corollary gives an
asymptotic $(1-\alpha)$ confidence interval for $\mu$ at the prediction horizon $t_0$.

\begin{cor}
\label{thm:mean}
Under Assumptions A.1--A.5, as $n\to\infty$, 
$$
\lim_{n\to\infty}\mathbb{P}\left(\mu\in \hat{\mu} \pm \left|g^{\prime}(\hat{\eta})\right|\sqrt{c(\alpha) \left(\sum_{k=1}^{p_n+L+1}\frac{v_k^{*2}}{\lambda_k}\right)}\right)\geq 1-\alpha,
$$
where $\hat{\eta}=\boldsymbol{V}^{\top}\hat{\boldsymbol{\gamma}}$, $v_k^*=\sum_{l=1}^{p_n+L+1} v_l e_{kl}$, $c(\alpha)=[p_n+L+1+\sqrt{2(p_n+L+1)} \Phi(1-\alpha)] / n$, and $\Phi(1-\alpha)$ is the $1-\alpha$ quantile of standard normal distribution.
\end{cor}

Corollary~\ref{thm:mean} follows from Theorem~\ref{thm:beta} and, together with
Theorem~\ref{thm:variance}(i), yields a feasible confidence interval by replacing
$\boldsymbol{\Theta}$ with its estimator $\hat{\boldsymbol{\Theta}}$.

Denote the eigenvectors and eigenvalues of the matrix $\left(\boldsymbol{\Theta}_{22}^{(-1)}\right)^{-1}$ by $\left(e_{1}^{1}, \lambda_{1}^{1}\right), \ldots,\left(e_{p_n}^{1}, \lambda_{p_n}^{1}\right)$, and let
$\boldsymbol{e}_k^{1}=\left(e_{k,1}^{1}, \ldots, e_{k, p_n}^{1}\right)^\top$ for $k=1, \ldots, p_n$. To quantify uncertainty in the functional coefficient, the following corollary provides
an asymptotic $(1-\alpha)$ simultaneous confidence band for $\beta(s)$.

\begin{cor}
\label{cor:beta} 
Let $\boldsymbol{\rho}(s)=\left(\rho_1(s),\rho_2(s),\ldots,\rho_{p_n}(s)\right)^{\top}$, $\beta(s)$ is estimated by $\hat{\beta}(s)=\hat{\boldsymbol{\beta}}^{\top}\boldsymbol{\rho}(s)$.
Under Assumptions A.1--A.3, A.6--A.8, as $n\to\infty$,
\begin{align}
    \label{eq:band_1}
\lim_{n\to\infty}\mathbb{P}\left(\beta(s)\in\hat{\beta}(s) \pm \sqrt{c_1(\alpha) \left(\sum_{k=1}^{p_n}\frac{\omega_k(s)^2}{\lambda_k^{1}}\right)}, \quad\mbox{for all $s\in[0,1]$}\right)\geq 1-\alpha,
\end{align}
where $\omega_k(s)=\sum_{l=1}^{p_n} \rho_l(s) e_{kl}^{1}$, $c_1(\alpha)=[p_n+\sqrt{2 p_n} \Phi(1-\alpha)] / n$. The confidence band in (\ref{eq:band_1}) is equivalent to 
\begin{align}
\label{eq:band_2}
\lim_{n\to\infty}\mathbb{P}\left(\beta(s)\in \hat{\beta}(s) \pm \sqrt{c_1(\alpha) \boldsymbol{\rho}(s)^{\top}\boldsymbol{\Theta}_{22}^{(-1)}\boldsymbol{\rho}(s)} , \quad\mbox{for all $s\in[0,1]$}\right)\geq 1-\alpha.
\end{align}
\end{cor}

Corollary~\ref{cor:beta} follows directly from Theorem~\ref{thm:scb_beta}.
The confidence band~\eqref{eq:band_1} is established using arguments similar to
Corollary~4.3 of \citet{muller2005generalized}, while~\eqref{eq:band_2} parallels
Corollary~1 of \citet{kim2024generalized}.
The representation~\eqref{eq:band_2} is computationally preferable, as it avoids explicit
eigendecomposition of $\left(\boldsymbol{\Theta}_{22}^{(-1)}\right)^{-1}$.

The following corollary concerns inference for the regression coefficients.
\begin{cor}
\label{cor:theta}
Assume that Assumptions~A.1--A.3 and~A.9 hold.

\medskip
\noindent
(i)\ \textit{Single-parameter confidence interval.}
For the $j$th component $\tilde{\theta}_j$ of $\tilde{\boldsymbol{\theta}}$,
$0\le j\le L$, a $100(1-\alpha)\%$ confidence interval is
\(
\hat{\tilde{\theta}}_j
\pm
z_{\alpha/2}
\sqrt{\boldsymbol{\Theta}_{11,jj}^{(-1)}/n},
\)
where $z_{\alpha/2}$ is the $1-\alpha/2$ quantile of the standard normal
distribution and $\boldsymbol{\Theta}_{11,jj}^{(-1)}$ is the $j$th diagonal
element of $\boldsymbol{\Theta}_{11}^{(-1)}$.

(ii)\ \textit{Joint confidence region.}
A $100(1-\alpha)\%$ joint confidence region for
$\tilde{\boldsymbol{\theta}}$ is given by
\[
\left\{
\boldsymbol{\theta} :
n(\hat{\tilde{\boldsymbol{\theta}}}-\boldsymbol{\theta})^{\top}
(\boldsymbol{\Theta}_{11}^{(-1)})^{-1}
(\hat{\tilde{\boldsymbol{\theta}}}-\boldsymbol{\theta})
\le
\chi^{2}_{L+1,\,1-\alpha}
\right\},
\]
where $\chi^{2}_{L+1,\,1-\alpha}$ is the $1-\alpha$ quantile of the chi-squared
distribution with $L+1$ degrees of freedom.
\end{cor}

Corollary~\ref{cor:theta} follows directly from Theorem~\ref{thm:scb_theta}.
In contrast to the confidence band for $\beta(s)$, which is conservative because it avoids
deriving the explicit limiting distribution of $\hat{\beta}(s)-\beta(s)$, the confidence
intervals and joint confidence regions for $\tilde{\boldsymbol{\theta}}$ are asymptotically exact,
as the limiting distribution of $\hat{\tilde{\boldsymbol{\theta}}}-\tilde{\boldsymbol{\theta}}$
is established in Theorem~\ref{thm:scb_theta}.

\section{Further Extensions}
\label{sec:4extensions}

The proposed methodology can be readily extended to the following two useful settings.

\subsection{Multiple functional covariates}
\label{sec:multiple}

Suppose that $Q$ functional predictors are observed. The linear predictor is
\(
\eta_i
=
\theta_0
+
\boldsymbol Z_i^\top \boldsymbol\theta
+
\sum_{q=1}^Q \int \beta_q(s)\{X_{iq}(s)-m_q(s)\}\,ds,
\)
where $X_{iq}(\cdot)\in\mathcal L^2[0,1]$ and
$m_q(s)=\mathbb E\{X_{iq}(s)\}$.

Using basis expansions,
\(
X_{iq}(s)
=
m_q(s)+\sum_{j=1}^\infty \varepsilon_{jq}^{(i)}\rho_{jq}(s),
\)
\(
\beta_q(s)
=
\sum_{j=1}^\infty \beta_{jq}\rho_{jq}(s),
\)
and truncating at $p_{q,n}$, the linear predictor reduces to
\(
\eta_i
=
\theta_0
+
\boldsymbol Z_i^\top\boldsymbol\theta
+
\sum_{q=1}^Q\sum_{j=1}^{p_{q,n}}
\beta_{jq}\varepsilon_{jq}^{(i)}.
\)
Let
\(
\boldsymbol\gamma
=
(\theta_0,\boldsymbol\theta^\top,\{\boldsymbol\beta_q^\top\}_{q=1}^Q)^\top
\)
and
\(
\boldsymbol V_i
=
(1,\boldsymbol Z_i^\top,\{\boldsymbol\varepsilon_{iq}^\top\}_{q=1}^Q)^\top,
\)
where
\(
\boldsymbol\varepsilon_{iq}
=
(\varepsilon_{1q}^{(i)},\ldots,
\varepsilon_{p_{q,n}q}^{(i)})^\top
\)
and 
\(
\boldsymbol\beta_q
=
(\beta_{1q},\ldots,\beta_{p_{q,n}q})^\top.
\)
The estimating equation retains the same form as \eqref{eq:GEE}.
The information matrix $\boldsymbol\Theta$ now has a $(1+Q)\times(1+Q)$ block
structure corresponding to the scalar and functional components.

Under regularity conditions analogous to those in 
Section~\ref{SEC:theory}, 
and by adapting the arguments of 
Theorem~\ref{thm:beta}, 
we obtain
\[
\frac{
n(\hat{\boldsymbol\gamma}-\boldsymbol\gamma)^\top
\boldsymbol\Theta
(\hat{\boldsymbol\gamma}-\boldsymbol\gamma)
-
\{1+L+\sum_{q=1}^Q p_{q,n}\}
}{
\sqrt{2\{1+L+\sum_{q=1}^Q p_{q,n}\}}
}
\xrightarrow{d}
N(0,1).
\]
Moreover, by an argument parallel to 
Theorem~\ref{thm:scb_beta}, 
for each $q=1,\ldots,Q$,
\[
\frac{
n(\hat{\boldsymbol\beta}_q-\boldsymbol\beta_q)^\top
(\boldsymbol\Theta_{q+1,q+1}^{(-1)})^{-1}
(\hat{\boldsymbol\beta}_q-\boldsymbol\beta_q)
-
p_{q,n}
}{
\sqrt{2p_{q,n}}
}
\xrightarrow{d}
N(0,1).
\]
Consequently, by Corollary~\ref{cor:beta}, 
an approximate $(1-\alpha)$ simultaneous confidence band for
$\beta_q(s)$ is given by
\[
\hat\beta_q(s)
\pm
\sqrt{
c_q(\alpha)\,
\boldsymbol\rho_q(s)^\top
\boldsymbol\Theta_{q+1,q+1}^{(-1)}
\boldsymbol\rho_q(s)
},
\]
where
$c_q(\alpha)
=
[p_{q,n}+\sqrt{2p_{q,n}}\Phi(1-\alpha)]/n$
and
$\boldsymbol\rho_q(s)
=
(\rho_{1q}(s),\ldots,\rho_{p_{q,n}q}(s))^\top$.

\subsection{Competing risks with missing causes of event}
\label{sec:competing}

In survival analysis with competing risks, the cause of an event may be
unknown due to incomplete data collection or recording issues
\citep{gao2005semiparametric, mao2017efficient}.
The proposed functional direct binomial method can be naturally extended
to accommodate missing causes of failure, as described below.

Assume two possible event types ($K=2$), indexed by $k=1,2$, with
type~1 being the event of interest.
Define the complete-case indicator
$R_i = I(D_i \in \{1,2\})$, where $R_i=1$ if the event type is observed
and $D_i \in \{1,2\}$ denotes the failure type.
Assume the cause of failure is missing at random (MAR), meaning that
\(
\mathbb P(R_i=1 \mid W_i, D_i)
=
\mathbb P(R_i=1 \mid W_i),
\)
where the observed data are
$W_i = (\tilde T_i, \boldsymbol Z_i, X_i(\cdot))$.
Let
\(
\pi(W_i) = \mathbb P(R_i=1 \mid W_i)
\)
denote the probability that the event type is observed,
and let $\hat\pi(W_i)$ be its estimator.

To adjust for the missing cause, we generalize \eqref{eq:GEE} as
\[
\frac{1}{n}
\sum_{i=1}^n
\frac{w_i}{\hat G(\tilde T_i \wedge t_0-)}
\frac{R_i}{\hat\pi(W_i)}
\frac{(\tilde Y_i-\mu_i) g'(\eta_i)}{\sigma^2(\mu_i)}
\boldsymbol V_i
=
0,
\]
where the additional term $R_i/\hat\pi(W_i)$ accounts for missing causes
through inverse probability weighting.
The asymptotic results are established under the additional assumption
that $R_i$ follows a logistic model, and the variability introduced by
estimating $\pi(\cdot)$ is incorporated into the variance formula;
see Appendix~A.10 for details.

\section{Simulation}
\label{SEC:simulation}

This section reports simulation results on estimation accuracy and confidence interval coverage of
the proposed functional direct binomial model under competing risks, for settings with one and
three functional covariates.

To ensure that the CIF at horizon $t_0$
satisfies model~\eqref{eq:model}, type~1 event times are generated from a
Fine--Gray subdistribution model \citep{fine1999proportional} with a log--log link.
Throughout the simulations, we set the prediction horizon to $t_0=5$.
We define
\[
F_1\{t_0\mid \boldsymbol Z_i=\mathbf 0_L, X_i(\cdot)\equiv 0\}
=
\pi\{1-\exp(-\lambda t_0^\nu)\},
\]
where
\(
F_1(t\mid \boldsymbol Z_i,X_i)
=
\mathbb P(T_i\le t, D_i=1 \mid \boldsymbol Z_i,X_i)
\)
denotes the CIF for a type~1 event, and
$\pi=\mathbb P(D_i=1\mid \boldsymbol Z_i=\mathbf 0_L, X_i(\cdot)\equiv 0)$
controls the baseline type~1 event rate.
According to the Fine--Gray model with log--log link,
\begin{equation}
\label{eq:simu}
\log\!\bigl[-\log\{1-F_1(t\mid \boldsymbol Z_i,X_i)\}\bigr]
=
\theta_0+\eta_{1,i},
\end{equation}
with intercept
$\theta_0=\log[-\log\{1-\pi(1-e^{-\lambda t_0^\nu})\}]$.
This construction is consistent with the data-generating mechanism in
\eqref{eq:model}, yielding
\(
F_1(t_0\mid \boldsymbol Z_i,X_i)
=
1-\bigl[1-\pi(1-e^{-\lambda t_0^\nu})\bigr]^{\exp(\eta_{1,i})}.
\)
Conditional on $D_i=1$, event times are generated from
\begin{equation}
\label{eq:con_1}
\mathbb P(T_i\le t_0\mid D_i=1,\boldsymbol Z_i,X_i)
=
\frac{
1-\bigl[1-\pi(1-e^{-\lambda t_0^\nu})\bigr]^{\exp(\eta_{1,i})}
}{
1-(1-\pi)^{\exp(\eta_{1,i})}
},
\end{equation}
using inverse transform sampling.
For the competing event ($D_i=2$), event times follow a Weibull distribution with
CIF
\begin{equation}
\label{eq:con_2}
\mathbb P(T_i\le t_0\mid D_i=2,\boldsymbol Z_i,X_i)
=
1-\exp\{-t_0\exp(\eta_{2,i})\}.
\end{equation}

Three settings are considered, each based on the sine basis
$\rho_j(s)=\sqrt{2}\sin(\pi j s)$ on $[0,1]$.

\textbf{Setting~1.}
Following \citet{muller2005generalized}, we generate
$X(s)=\sum_{j=1}^{20}\varepsilon_j\rho_j(s)$ with independent
$\varepsilon_j\sim N(0,1/j^2)$.
The coefficient function is
$\beta(s)=\sum_{j=1}^{20}\beta_j\rho_j(s)$, where $\beta_j=1/j$ for $j\le3$
and $\beta_j=0$ otherwise.

\textbf{Setting~2.}
We generate $X_k(s)=\sum_{j=1}^{20}\varepsilon_{jk}\rho_j(s)$ for $k=1,2,3$,
with
$\varepsilon_{j1}\sim N(0,1/j^2)$,
$\varepsilon_{j2}\sim N(0,1/j^{1/2})$, and
$\varepsilon_{j3}\sim N(0,1)$.
The corresponding coefficient functions
$\beta_k(s)=\sum_{j=1}^{20}\beta_{jk}\rho_j(s)$ satisfy
$\beta_{j1}=1/j$ for $j\le3$,
$\beta_{j2}=0.5$ for $j=1,2$,
$\beta_{j3}=1$ for $j=1$,
and $\beta_{jk}=0$ otherwise.

{
In Settings~1 and~2, we use working sieve dimensions
$p_n=3$ and $(p_{n,1},p_{n,2},p_{n,3})=(3,2,1)$, respectively,
matching the finite nonzero basis representations of the coefficient functions.

\textbf{Setting~3.}
To examine estimation when truncation leaves a nonzero residual component,
we retain the functional-predictor generation mechanism of Setting~1,
\(
X(s)=\sum_{j=1}^{20}\varepsilon_j\rho_j(s)
\)
with
\(
\varepsilon_j\sim N(0,1/j^2),
\)
but use the coefficient function
\(
\beta(s)=\sum_{j=1}^{20}j^{-2}\rho_j(s),
\)
so that all 20 basis coefficients are nonzero. The primary simulation uses
the working sieve dimension $p_n=5$, which retains approximately $91.7\%$
of the total variance of $X(\cdot)$, calculated as
\(
\sum_{j=1}^{5}j^{-2}/\sum_{j=1}^{20}j^{-2}.
\)
Hence,
\(
X(s)
=
\sum_{j=1}^{p_n}\varepsilon_j\rho_j(s)
+
\sum_{j>p_n}^{20}\varepsilon_j\rho_j(s),
\)
and the second term remains nonzero after truncation. Because
$\beta_j=j^{-2}$ is also nonzero for $j>p_n$, the omitted part contributes
\(
\sum_{j>p_n}^{20}\beta_j\varepsilon_j
\)
to the functional linear predictor. Since
$\operatorname{Var}(\varepsilon_j)=j^{-2}$, the variance of this omitted
contribution is
\(
\sum_{j>p_n}^{20}j^{-6},
\)
which decreases rapidly with $p_n$.

To assess sensitivity to the sieve dimension, we additionally repeat
Setting~3 with the more restrictive choice $p_n=2$, which retains
approximately $78.3\%$ of the total variance of $X(\cdot)$. All pointwise
estimation and inference results for $\beta(s)$ are evaluated against the
full 20-term coefficient function rather than its projection onto the
working sieve. Unlike Settings~1 and~2, Setting~3 therefore retains a nonzero
truncation residual in the functional linear predictor.
}

Three scalar covariates are generated as
$Z_1\sim\mathrm{Binom}(1,0.75)$,
$Z_2\sim\mathrm{Binom}(1,0.1)$, and
$Z_3\sim N(0,1)$, with coefficients
$\boldsymbol\theta=(0.5,-0.2,0.2)^\top$.
The linear predictors in \eqref{eq:simu} and \eqref{eq:con_2} are
$\eta_{1,i}=\boldsymbol Z_i^\top\boldsymbol\theta
+\sum_{k=1}^K\int\beta_k(s)X_{ik}(s)\,ds$
and
$\eta_{2,i}=\boldsymbol Z_i^\top(-\boldsymbol\theta)
+\sum_{k=1}^K\int\{-\beta_k(s)\}X_{ik}(s)\,ds$.

Let $\pi=0.9$.
The event type $D_i$ is generated independently as
$1+\mathrm{Bernoulli}\{(1-\pi)^{\exp(\eta_{1,i})}\}$, where $D_i=1$ denotes the
event of interest and $D_i=2$ a competing event.
Event times are generated from the conditional distributions
\eqref{eq:con_1} or \eqref{eq:con_2} via inverse transform sampling, depending on
$D_i$, following \citet{kawaguchi2019fast}.
Censoring times are generated independently from an exponential distribution
with scale $0.05$, and the observed time is the minimum of the event and censoring
times.
The Weibull shape parameter is fixed at $\nu=1$, with scale
$\lambda\in\{0.02,0.1,0.3\}$.
Sample sizes are $n\in\{500,1000,5000\}$, and each setting is replicated $R=500$
times.

The average censoring rates under Settings~1 and~2 are summarized in
Table~S2 in Appendix~C.
When $\lambda=0.3$, the censoring rate is the lowest, resulting in minimal
information loss.
When $\lambda=0.02$, the censoring rate reaches approximately $0.711$ under
Setting~1, closely mimicking the real data used in Section~\ref{SEC:app}.

Table~\ref{table:mse} evaluates the estimation accuracy of the proposed
functional direct binomial model under varying censoring rates and sample sizes
in Setting~1 and additionally reports pointwise results for $\beta(s)$ at the
representative locations $s=0.25$, $0.50$, and $0.75$. The corresponding
coefficient-level results for Setting~2 with multiple functional predictors
are reported in Table~S3, and the pointwise functional-coefficient
results are reported in Table~S4, in
Appendix~C. For Setting~3,
Table~S5 reports results under the primary sieve dimension
$p_n=5$, while Table~S6 reports the corresponding
sensitivity results for $p_n=2$.

For a scalar parameter such as the intercept $\theta_0$, the Mean Squared Error
(MSE), Bias, and estimated Standard Error (SE$_{\mathrm{est}}$) are computed as
\(
\mathrm{MSE}(\hat\theta_0)
=
\frac{1}{R}\sum_{r=1}^{R}(\hat\theta_{0,r}-\theta_0)^2
\),
\(
\mathrm{Bias}(\hat\theta_0)
=
\left|
\frac{1}{R}\sum_{r=1}^{R}\hat\theta_{0,r}-\theta_0
\right|
\),
and
\(
\mathrm{SE}_{\mathrm{est}}(\hat\theta_0)
=
\sqrt{
\frac{1}{R}\sum_{r=1}^{R}
\left(
\hat\theta_{0,r}
-
\frac{1}{R}\sum_{r'=1}^{R}\hat\theta_{0,r'}
\right)^2
},
\)
where $\hat\theta_{0,r}$ denotes the estimate from the $r$-th replication.
The same definitions of MSE, Bias, and empirical
SE$_{\mathrm{est}}$ are applied to $\hat\beta(s)$ at the three representative
locations.

Estimation accuracy generally improves with increasing sample size in
Settings~1 and~2. In particular, larger sample sizes lead to smaller MSE and
lower empirical variability, while the absolute biases remain generally small.
Moreover, lower censoring rates generally result in better estimation accuracy,
reflecting reduced information loss. For the functional coefficient in
Setting~1, the pointwise MSE and empirical variability at $s=0.25$, $0.50$,
and $0.75$ decrease markedly as the sample size increases. Similar trends are
observed for the three functional coefficients in Setting~2, as reported in
Table~S4.

To assess interval validity, Tables~\ref{table:mse},
S3, S4, S5, and
S6 also report empirical coverage probabilities (CR),
defined as the proportion of replications in which the true parameter value is
contained within the corresponding confidence interval or band. The $95\%$
confidence intervals for the scalar parameters $\theta_0$, $\theta_1$,
$\theta_2$, and $\theta_3$ are constructed based on
Corollary~\ref{cor:theta}. For the functional coefficients, we additionally
report pointwise $95\%$ confidence intervals at $s=0.25$, $0.50$, and $0.75$,
which are constructed using the pointwise asymptotic variance of
$\hat\beta(s)$, while the $95\%$ simultaneous confidence bands for the entire
coefficient functions are based on Corollary~\ref{cor:beta}. In Setting~1,
the pointwise coverage probabilities at the three representative locations
range from 0.916 to 0.960 across the simulation configurations, while the
simultaneous band coverage remains close to the nominal level. The
corresponding pointwise coverage probabilities in Setting~2 range from 0.910
to 0.960, with similar behavior for the simultaneous bands. For the scalar
parameters and simultaneous confidence bands, the coverage probabilities
remain generally close to the nominal $95\%$ level in Settings~1 and~2.
Even under the configuration with the highest censoring rate and the smallest
sample size, the scalar parameters and functional coefficient achieve
near-nominal coverage.

{
Setting~3 examines estimation and inference for the functional coefficient
when a nonzero truncation residual remains. With the primary choice
$p_n=5$, the pointwise MSE and empirical variability decrease with
increasing sample size, while the pointwise coverage probabilities at
$s=0.25$, $0.50$, and $0.75$ remain close to the nominal level. The
simultaneous confidence-band coverage is also stable across the simulation
configurations.

The sensitivity analysis with $p_n=2$ shows the effect of a more restrictive
truncation. Although the smaller sieve dimension reduces empirical variability
and can yield smaller pointwise MSE in some configurations, non-negligible
bias remains at some locations. As the sample size increases, pointwise
coverage deteriorates at $s=0.50$ and $s=0.75$, and simultaneous
confidence-band coverage becomes poor. In contrast, $p_n=5$ maintains stable
pointwise and simultaneous coverage in the configurations considered.
}

\begin{figure}[h]
  \centering
  \includegraphics[width=0.9\textwidth]{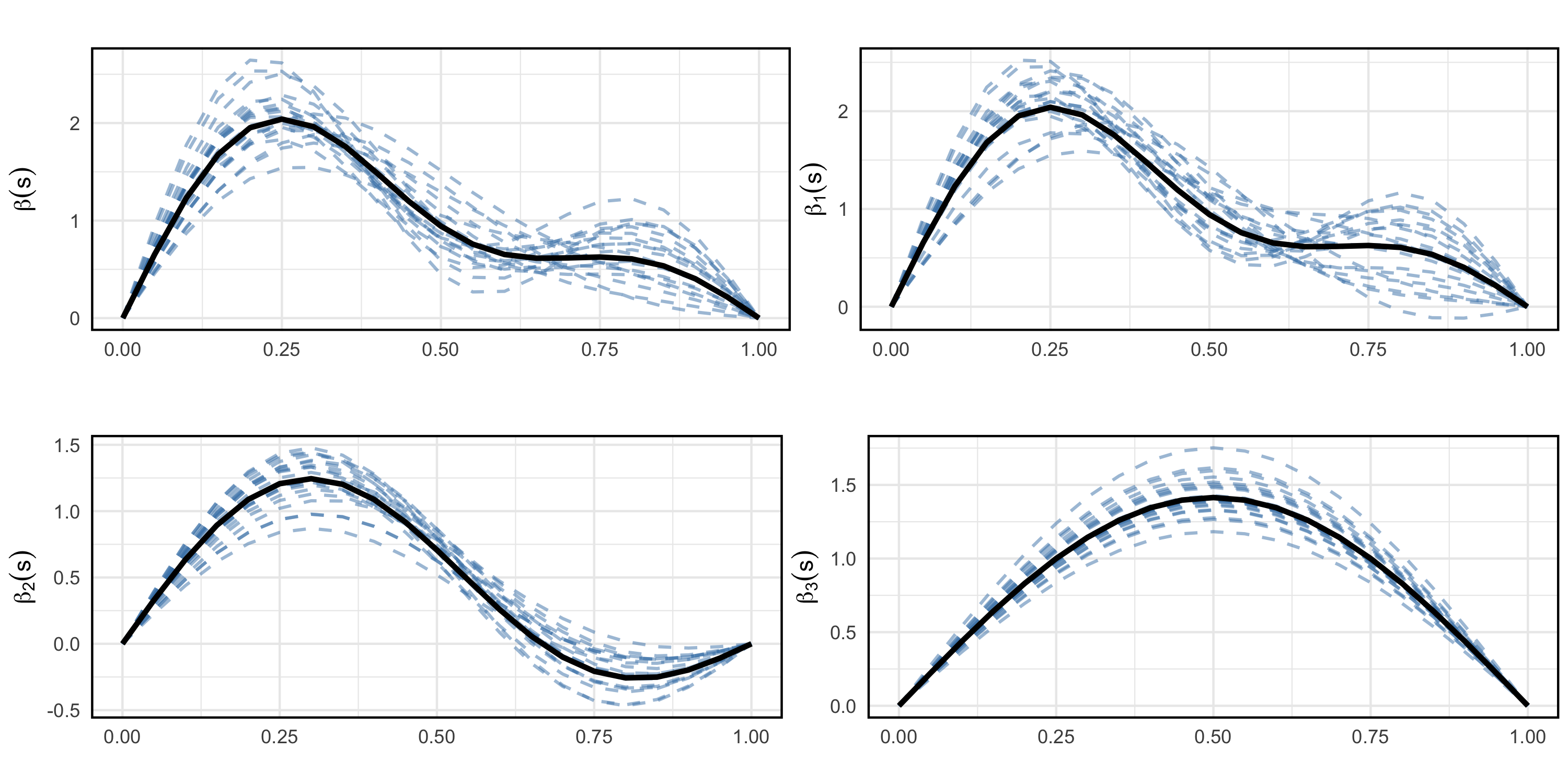}
  \caption{True functional coefficients $\beta(s)$ (solid line) and the estimated
  functions from 20 randomly selected Monte Carlo replications (dashed lines)
  for $\lambda=0.1$ and $n=1000$. The top-left panel corresponds to
  Setting~1; the remaining panels display the three coefficients under
  Setting~2.}
  \label{fig:sim_est}
\end{figure}

Figure~\ref{fig:sim_est} displays the true function $\beta(s)$ together with
estimated functions from the Monte Carlo replications
under Settings~1 and~2.
The estimated curves cluster tightly around the truth across the domain,
indicating accurate recovery of the underlying functional structure and
limited sampling variability in finite samples.

All code is publicly available at
{\tt https://github.com/jonathanhori/Censored-functional-GLM.}

\begin{table}
\centering
\caption{MSE, Bias, SE$_{est}$, and CR for different event rates under Setting~1.
For $\beta(0.25)$, $\beta(0.50)$, and $\beta(0.75)$,
CR denotes the coverage probability of the pointwise 95\% confidence interval;
for $\beta(s)$ (SCB), CR denotes simultaneous 95\% confidence-band coverage.}
\label{table:mse}
\renewcommand{\arraystretch}{0.55}
\resizebox{\linewidth}{!}
{\begin{tabular}{cccccccccccccc}
\toprule
& & \multicolumn{4}{c}{$\lambda=0.02$}
& \multicolumn{4}{c}{$\lambda=0.1$}
& \multicolumn{4}{c}{$\lambda=0.3$} \\
\cmidrule(l{3pt}r{3pt}){3-6}
\cmidrule(l{3pt}r{3pt}){7-10}
\cmidrule(l{3pt}r{3pt}){11-14}
& $n$ & MSE & Bias & SE$_{est}$ & CR
& MSE & Bias & SE$_{est}$ & CR
& MSE & Bias & SE$_{est}$ & CR \\
\midrule

\multirow{3}{*}{$\theta_0$}
& 500  & 0.103 & 0.053 & 0.316 & 0.950
       & 0.034 & 0.009 & 0.184 & 0.950
       & 0.029 & 0.021 & 0.169 & 0.914\\
& 1000 & 0.055 & 0.038 & 0.232 & 0.932
       & 0.016 & 0.005 & 0.127 & 0.950
       & 0.011 & 0.001 & 0.106 & 0.922\\
& 5000 & 0.009 & 0.008 & 0.093 & 0.962
       & 0.003 & 0.001 & 0.053 & 0.952
       & 0.002 & 0.001 & 0.049 & 0.954\\
\midrule

\multirow{3}{*}{$\theta_1$}
& 500  & 0.103 & 0.017 & 0.321 & 0.956
       & 0.039 & 0.008 & 0.197 & 0.936
       & 0.035 & 0.004 & 0.187 & 0.930\\
& 1000 & 0.056 & 0.015 & 0.235 & 0.932
       & 0.020 & 0.004 & 0.142 & 0.938
       & 0.016 & 0.011 & 0.128 & 0.934\\
& 5000 & 0.009 & 0.007 & 0.097 & 0.966
       & 0.003 & 0.001 & 0.058 & 0.960
       & 0.003 & 0.003 & 0.054 & 0.934\\
\midrule

\multirow{3}{*}{$\theta_2$}
& 500  & 0.226 & 0.081 & 0.468 & 0.962
       & 0.079 & 0.020 & 0.280 & 0.954
       & 0.075 & 0.013 & 0.274 & 0.942\\
& 1000 & 0.099 & 0.011 & 0.314 & 0.950
       & 0.035 & 0.008 & 0.186 & 0.960
       & 0.029 & 0.016 & 0.171 & 0.954\\
& 5000 & 0.018 & 0.003 & 0.134 & 0.948
       & 0.007 & 0.003 & 0.084 & 0.940
       & 0.006 & 0.002 & 0.078 & 0.952\\
\midrule

\multirow{3}{*}{$\theta_3$}
& 500  & 0.015 & 0.010 & 0.122 & 0.956
       & 0.007 & 0.003 & 0.085 & 0.930
       & 0.008 & 0.009 & 0.089 & 0.940\\
& 1000 & 0.007 & 0.003 & 0.086 & 0.946
       & 0.004 & 0.003 & 0.061 & 0.914
       & 0.003 & 0.003 & 0.057 & 0.940\\
& 5000 & 0.001 & 0.002 & 0.037 & 0.968
       & 0.001 & 0.000 & 0.026 & 0.948
       & 0.001 & 0.001 & 0.024 & 0.954\\
\midrule

\multirow{3}{*}{$\beta_1$}
& 500  & 0.021 & 0.028 & 0.141 & --
       & 0.013 & 0.023 & 0.112 & --
       & 0.016 & 0.034 & 0.122 & --\\
& 1000 & 0.010 & 0.013 & 0.102 & --
       & 0.006 & 0.013 & 0.074 & --
       & 0.007 & 0.016 & 0.080 & --\\
& 5000 & 0.002 & 0.002 & 0.041 & --
       & 0.001 & 0.004 & 0.033 & --
       & 0.001 & 0.003 & 0.035 & --\\
\midrule

\multirow{3}{*}{$\beta_2$}
& 500  & 0.071 & 0.019 & 0.266 & --
       & 0.031 & 0.006 & 0.176 & --
       & 0.032 & 0.028 & 0.176 & --\\
& 1000 & 0.030 & 0.004 & 0.172 & --
       & 0.014 & 0.004 & 0.116 & --
       & 0.013 & 0.011 & 0.113 & --\\
& 5000 & 0.006 & 0.000 & 0.078 & --
       & 0.002 & 0.001 & 0.048 & --
       & 0.003 & 0.004 & 0.050 & --\\
\midrule

\multirow{3}{*}{$\beta_3$}
& 500  & 0.138 & 0.018 & 0.372 & --
       & 0.060 & 0.009 & 0.245 & --
       & 0.066 & 0.026 & 0.255 & --\\
& 1000 & 0.069 & 0.014 & 0.262 & --
       & 0.027 & 0.012 & 0.163 & --
       & 0.029 & 0.000 & 0.170 & --\\
& 5000 & 0.013 & 0.002 & 0.115 & --
       & 0.007 & 0.000 & 0.081 & --
       & 0.005 & 0.004 & 0.071 & --\\
\midrule

\multirow{3}{*}{$\beta(0.25)$}
& 500  & 0.304 & 0.072 & 0.547 & 0.948
       & 0.150 & 0.040 & 0.386 & 0.930
       & 0.176 & 0.099 & 0.408 & 0.932\\
& 1000 & 0.146 & 0.034 & 0.380 & 0.944
       & 0.069 & 0.031 & 0.260 & 0.956
       & 0.073 & 0.031 & 0.267 & 0.938\\
& 5000 & 0.028 & 0.004 & 0.168 & 0.932
       & 0.013 & 0.005 & 0.114 & 0.942
       & 0.014 & 0.004 & 0.117 & 0.938\\
\midrule

\multirow{3}{*}{$\beta(0.50)$}
& 500  & 0.294 & 0.015 & 0.542 & 0.950
       & 0.141 & 0.020 & 0.375 & 0.950
       & 0.152 & 0.010 & 0.390 & 0.916\\
& 1000 & 0.156 & 0.002 & 0.394 & 0.944
       & 0.061 & 0.002 & 0.247 & 0.948
       & 0.063 & 0.024 & 0.250 & 0.938\\
& 5000 & 0.030 & 0.001 & 0.173 & 0.950
       & 0.015 & 0.006 & 0.121 & 0.930
       & 0.012 & 0.009 & 0.110 & 0.950\\
\midrule

\multirow{3}{*}{$\beta(0.75)$}
& 500  & 0.322 & 0.019 & 0.567 & 0.922
       & 0.125 & 0.023 & 0.353 & 0.932
       & 0.124 & 0.021 & 0.352 & 0.932\\
& 1000 & 0.134 & 0.022 & 0.365 & 0.956
       & 0.053 & 0.019 & 0.230 & 0.960
       & 0.058 & 0.001 & 0.241 & 0.922\\
& 5000 & 0.026 & 0.005 & 0.161 & 0.950
       & 0.012 & 0.003 & 0.109 & 0.932
       & 0.010 & 0.007 & 0.098 & 0.954\\
\midrule

\multirow{3}{*}{$\beta(s)$ (SCB)}
& 500  & -- & -- & -- & 0.946
       & -- & -- & -- & 0.932
       & -- & -- & -- & 0.942\\
& 1000 & -- & -- & -- & 0.960
       & -- & -- & -- & 0.952
       & -- & -- & -- & 0.958\\
& 5000 & -- & -- & -- & 0.950
       & -- & -- & -- & 0.944
       & -- & -- & -- & 0.942\\

\bottomrule
\end{tabular}}
\end{table}

\section{Real Data Example}
\label{SEC:app}

We illustrate the proposed method using CGM data to address two questions:  
(i) whether CGM-derived functional trajectories are significantly associated with advanced chronic kidney disease (CKD) risk after accounting for demographic and clinical covariates, and  
(ii) whether CGM-based functional measures can serve as a viable alternative to HbA1c, the conventional biomarker for diabetes management.


The study linked Dexcom CGM data with VA electronic health records \citep{okuno2024assessing}. Eligible patients were Veterans aged 21 years or older with diabetes whose first Dexcom sensor prescription occurred between 2015 and 2020. Daily CGM metrics were collected during a 183-day landmark period (six months), and patients were followed for up to five years for incident CKD stage~3a or higher (defined in Appendix D), with all-cause mortality treated as a competing risk. Accordingly, we set the prediction horizon to $t_0=5$ years in the real-data analysis. This landmark period length was chosen because of the strong association of CGM summary metrics with all-cause mortality in a six-month period, compared with shorter 14-day or three-month periods explored in \citet{okuno_continuous_2025}. After excluding patients with terminal events during the landmark period and those with prior CKD, 1,595 patients remained. Within five years of CGM initiation, 14\% of subjects developed CKD stage 3a or higher and 5.8\% died without CKD, yielding an 80\% censoring rate (Table~S7 in Appendix C).
 We evaluated three sets of candidate predictors: (i) occupation-time curves of CGM metrics (defined in Appendix E), including estimated mean glucose value (EGV), time in range (TIR), time above range (TAR), time below range (TBR), and coefficient of variation (CV); (ii) routinely collected clinical features, including average HbA1c over the landmark period; and (iii) demographic variables. Model performance was assessed using time-dependent AUC \citep{heagerty2000time} and Brier score \citep{graf1999assessment}.


To evaluate the contribution of CGM trajectories, we applied the proposed estimation framework to choose functional predictors, select their truncation parameters, and identify clinical features. The dataset was randomly partitioned into training (60\%) and validation (40\%) subsets. 
Within the training set, we considered candidate truncation
levels $p_n\in\{1,\ldots,10\}$ and selected $p_n$ using 10-fold
cross-validation based on time-dependent AUC with a sine basis.
The full selection procedure is described in
Section~F of the Appendix.
Using this procedure, we evaluated cross-validation predictive performance across all metrics and their pairwise combinations. 
This process identified EGV and TBR as the most informative functional predictors.  
For the final model, the selected truncation levels were
$p_{n,EGV}=3$ and $p_{n,TBR}=2$.
These predictors were combined with age at index as baseline covariates. A forward feature selection algorithm was then used to identify additional predictors that improved validation set AUC, resulting in the inclusion of baseline creatinine and LDL, ethnicity, hypertension status, Medicare status, and sex.  The final model results are summarized in Table~\ref{table:effect_sizes_ckd_final}. 

\begin{figure}
\includegraphics[width=0.9\linewidth,height=0.3\textheight]{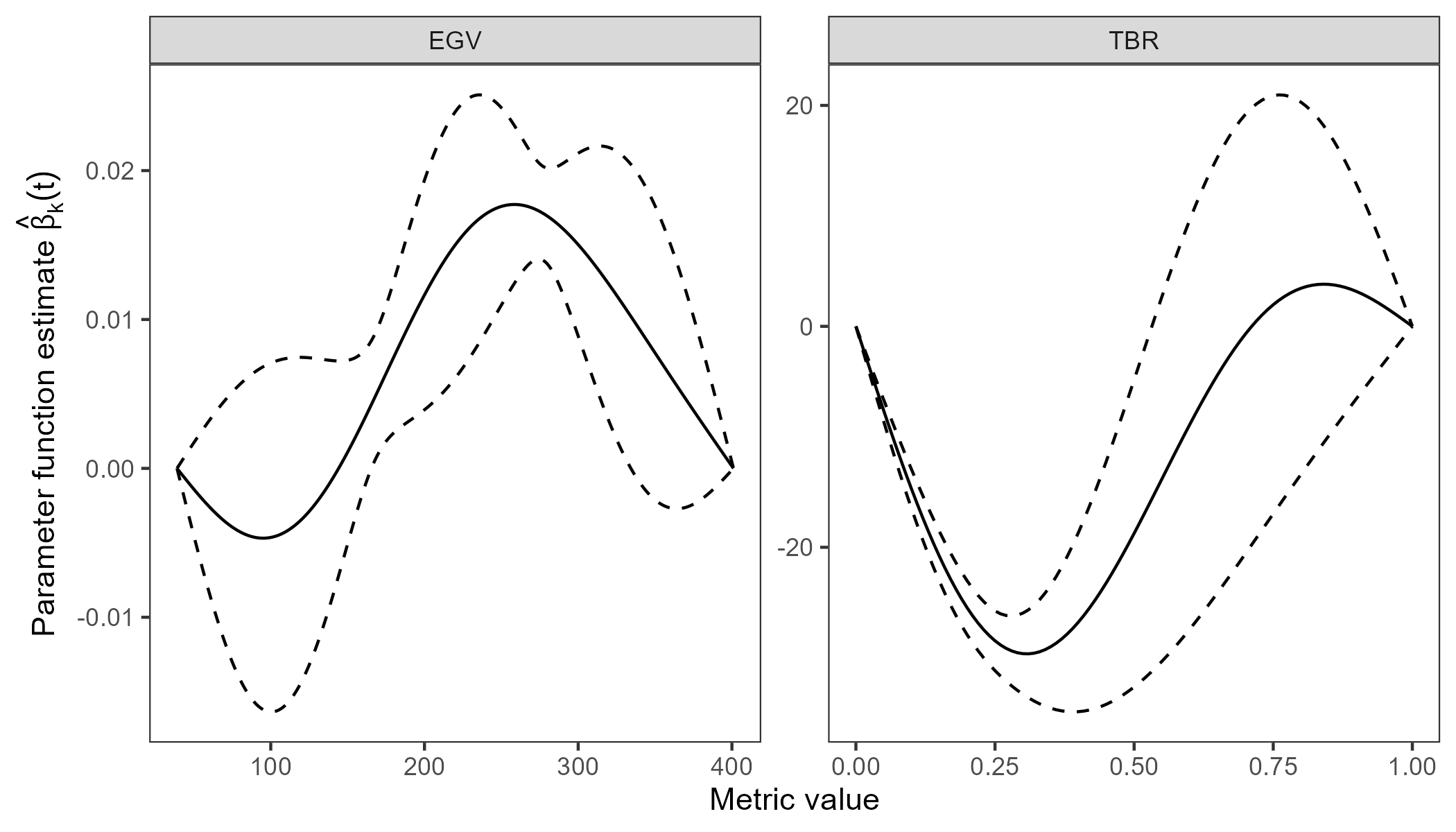}
\caption{Estimated parameter functions $\hat\beta_k(s)$ with confidence bands
in CKD model controlling for selected scalar covariates. The truncation
parameters maximizing the cross-validation AUC within the training set was
$p_{n, EGV} = 3$ and $p_{n, TBR} = 2$.}
\label{figure:betahat_ckd}
\end{figure}

It is seen from Table~\ref{table:effect_sizes_ckd_final} that 
the functional predictors EGV and TBR are highly significant risk factors for CKD. In particular, the individual tests of $H_0:\beta_k(s)=0$ for $k=1$ (EGV) and $k=2$ (TBR) each yielded $p<0.001$, and the confidence bands of the estimated functions \(\hat\beta_k(s)\), displayed in Figure~\ref{figure:betahat_ckd}, exclude zero over substantial regions. These intervals indicate CGM metric thresholds above which a longer time in the landmark period is significantly associated with outcome risk. For example, for any value $u\in [200, 300]$, patients having more landmark days with EGV readings above $u$ have an increased risk for CKD.
We also observe that TBR and LDL appear to exhibit protective effects on CKD risk, which is counterintuitive. This pattern may reflect selection bias, as individuals with elevated TBR or LDL are more likely to die earlier, leaving a survivor group with seemingly lower CKD risk—a phenomenon well documented in the competing risks literature \citep{young2020causal, hernan2010hazards, mansournia2022practical}.

\begin{table}[h!]
\centering
\caption{Estimated coefficients with 95\% confidence intervals and $p$-values in the final CKD risk prediction model.}
\renewcommand{\arraystretch}{0.55}
\begin{tabular}{lccc}
\toprule
Feature & Estimate & 95\% CI & $p$-value \\
\midrule

Intercept & -3.115 & {}[-3.281, -2.949] & $<$0.001\\

\addlinespace[0.8ex]
\multicolumn{4}{l}{\textbf{Scalar covariates}} \\
\addlinespace[0.5ex]


Age at index & 0.434 & {}[0.354, 0.514] & $<$0.001\\
Male & 0.097 & {}[-0.037, 0.231] & 0.157\\
Ethnicity (vs.\ White Non-Hispanic) & & & \\
\hspace{1em}African American & 1.144 & {}[0.904, 1.384] & $<$0.001\\
\hspace{1em}White Hispanic & 0.690 & {}[0.301, 1.080] & $<$0.001\\
\hspace{1em}Other Ethnicity & 0.256 & {}[-0.065, 0.576] & 0.118\\
Has Medicare & 0.225 & {}[0.062, 0.389] & 0.007\\
Hypertensive & 0.865 & {}[0.734, 0.997] & $<$0.001\\
Creatinine & 0.861 & {}[0.789, 0.933] & $<$0.001\\
LDL & -0.179 & {}[-0.237, -0.121] & $<$0.001\\

\addlinespace[0.8ex]
\multicolumn{4}{l}{\textbf{Functional covariates}} \\
\addlinespace[0.5ex]


EGV ($p_{n, EGV}=3$)& \multicolumn{2}{c}{See Fig.~2} & $<$0.001\\
TBR ($p_{n, TBR}=2$)& \multicolumn{2}{c}{See Fig.~2} & $<$0.001\\

\bottomrule
\end{tabular}
\label{table:effect_sizes_ckd_final}
\end{table}

\begin{figure}
\includegraphics[width=0.9\linewidth,height=0.3\textheight]{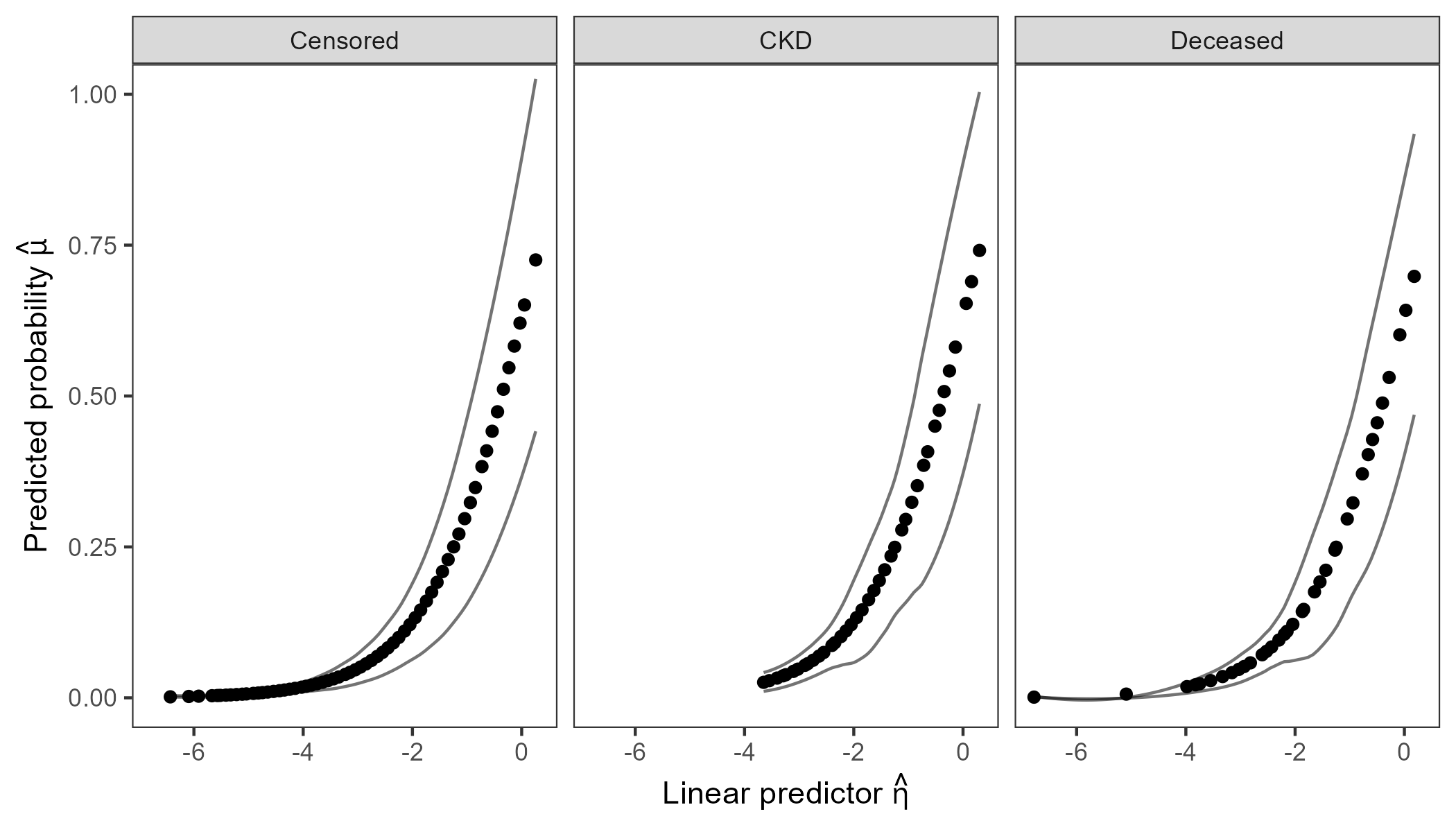}
\caption{Linear predictor $\hat\eta$ versus predicted probability $\hat\mu$ with 95\% confidence intervals for selected patients. The model includes selected
scalar and functional covariates ($p_{n, EGV}=3$, $p_{n, TBR} = 2$). Confidence intervals are smoothed for display.}
\label{figure:linear_predictor_ckd}
\end{figure}

In addition, we visualize a sample of the subjects’ linear predictors $\hat\eta$ versus their predicted probabilities $\hat\mu$ and 95\% confidence intervals in Figure~\ref{figure:linear_predictor_ckd}. This figure illustrates that our proposed framework enables estimation of individualized predicted probabilities for CKD, along with associated confidence intervals.

We next compared the predictive performance of CGM- and HbA1c-based models using 10-fold cross-validation on the full dataset. As shown in Table~\ref{table:results_occ_compiled_ckd}, the CGM-based model achieved performance nearly identical to that of the HbA1c-based model (AUC 0.748 vs.\ 0.744; Brier 0.121 vs.\ 0.120). 
These findings suggest that CGM-derived measures offer predictive performance for CKD comparable to that of HbA1c, while providing the practical benefit of continuous, device-based monitoring rather than intermittent in-person blood testing.

\begin{table}[h!]
\centering
\caption{Predictive performance of competing CKD risk models (CGM vs.\ HbA1c) from 10-fold cross validation. Standard errors in parentheses are calculated using the per-fold evaluation metrics.}
\renewcommand{\arraystretch}{0.55}
\begin{tabular}{lcc}
\toprule
Model & AUC (SE) & Brier (SE) \\
\midrule

HbA1c + Selected scalars & 0.744 (0.061) & 0.120 (0.026)\\
CGM + Selected scalars & 0.748 (0.069) & 0.121 (0.035)\\
\bottomrule
\end{tabular}
\label{table:results_occ_compiled_ckd}
\end{table}

\section{Discussion}
\label{sec:discussion}

In this work, we developed a horizon-specific direct binomial regression framework
for right-censored competing risks data with scalar and dense functional predictors.
By directly modeling the cause-specific CIF, the proposed approach avoids
proportional hazards assumptions while remaining computationally simple and
interpretable. The framework accommodates multiple functional covariates and
can be extended to settings with missing event causes. To our knowledge, it is
the first method that enables direct modeling of cause-specific CIFs with dense
functional data, together with joint asymptotic inference for scalar, basis, and
functional coefficients.

We illustrated the proposed method using CGM data
to predict CKD risk among Veterans with diabetes, treating
death as a competing event. Linking CGM records with electronic health data, we
found that CGM-derived functional measures are significantly associated with CKD
risk and achieve predictive performance comparable to landmark-period HbA1c,
supporting functional CGM data as a valuable alternative prognostic source for
modeling diabetes complications.

Although the horizon-specific formulation allows regression effects to vary across
prediction horizons, such flexibility may reduce estimation efficiency, particularly
under heavy censoring and low event rates, as observed in the CKD application.
Ongoing work therefore considers extensions of Fine--Gray and transformation models
with functional covariates, which may offer improved efficiency in highly censored
settings.

An additional consideration arises when interest lies in modeling
the CIFs of multiple competing event types simultaneously. Because the proposed
framework models each cause-specific CIF marginally, separately fitted models do not
automatically enforce the constraint that the sum of the predicted CIFs across causes
is no greater than one. The present work therefore focuses on marginal estimation and
inference for a prespecified cause. Simultaneous prediction of multiple competing
events would require a joint or constrained modeling framework that explicitly
enforces coherence among the cause-specific CIFs. A full treatment of joint
multi-cause modeling with functional covariates is beyond the scope of this paper
and is left for future work.

More broadly, modern biobank studies increasingly collect multi-modal and
high-dimensional measurements, motivating extensions of the framework to
high-dimensional scalar covariates. In our application, CGM information is
summarized using daily aggregated metrics, where long landmarking periods and
strong temporal collinearity complicate interpretation of $\hat\beta(s)$ and may
discard informative intra-day glucose dynamics. Extensions to high-frequency
CGM data, such as occupation-time representations, may improve both predictive
performance and interpretability of functional effects.

\begin{acknowledgement}
\label{funding}

This work was supported by National Institutes of Health under grants
P30 CA-16042 (GL), UL1TR000124-02 (GL), T32 HG002536 (JH),
R35 GM141798 (HZ), and R01 DK142026 (HZ, JZ and GL);
National Science Foundation under grants DMS 2054253 (HZ and JZ)
and IIS 2205441 (SZ, JH, HZ, JZ and GL).
\end{acknowledgement}
\vspace*{1pc}

\noindent {\bf{Conflict of Interest}}

\noindent {\it{The authors have declared no conflict of interest.}}

\section*{Appendix}
The Supplementary Material contains detailed theoretical proofs and assumptions
(Section~A), simplified score functions
(Section~B), additional simulation tables
(Section~C), the outcome definition
(Section~D), occupation time curves
(Section~E), and the procedure for selecting the
truncation level $p_n$ (Section~F).

\end{document}